\documentclass{nature}

\usepackage{cite}
\usepackage[dvipdfmx]{graphicx}
\usepackage{lineno}
\usepackage{bm}
\usepackage{color}
\usepackage{amsmath}
\usepackage{amssymb}
\usepackage{empheq}
\usepackage{float}
\usepackage{multirow}
\usepackage{cases}
\usepackage{mathrsfs}
\usepackage{braket}

\newcommand{\onlinecite}[1]{\hspace{-1 ex} \nocite{#1}\citenum{#1}}

\title{Tunable spin-valley coupling in layered polar Dirac metals}

\author{Masaki Kondo$^{1, *}$, Masayuki Ochi$^1$, Tatsuhiro Kojima$^2$, Ryosuke Kurihara$^3$, Daiki Sekine$^4$, Masakazu Matsubara$^{4, 5}$, Atsushi Miyake$^3$, Masashi Tokunaga$^3$, Kazuhiko Kuroki$^1$, Hiroshi Murakawa$^1$, \\Noriaki Hanasaki$^1$, and Hideaki Sakai$^{1, 6, \dagger}$}

\begin{document}

\maketitle

\begin{affiliations}
	\item Department of Physics, Osaka University, Toyonaka, Osaka 560-0043, Japan
	\item Department of Chemistry, Osaka University, Toyonaka, Osaka 560-0043, Japan
	\item The Institute for Solid State Physics, The University of Tokyo, Kashiwa, Chiba 277-8581, Japan
	\item Department of Physics, Tohoku University, Sendai 980-8578 Japan
	\item Center for Science and Innovation in Spintronics, Tohoku Unviersity, Sendai 980-8577, Japan
	\item PRESTO, Japan Science and Technology Agency, Kawaguchi, Saitama 332-0012, Japan
\end{affiliations}

E-mail: $^*$kondo@gmr.phys.sci.osaka-u.ac.jp, $^{\dagger}$sakai@phys.sci.osaka-u.ac.jp.


\section*{Abstract}
\begin{abstract}
In non-centrosymmetric metals, spin-orbit coupling induces momentum-dependent spin polarization at the Fermi surfaces. 
This is exemplified by the valley-contrasting spin polarization in monolayer transition metal dichalcogenides with in-plane inversion asymmetry.
However, the valley configuration of massive Dirac fermions in transition metal dichalcogenides is fixed by the graphene-like structure, which limits the variety of spin-valley coupling.
Here, we show that the layered polar metal BaMn$X_2$ ($X=$Bi, Sb) hosts tunable spin-valley-coupled Dirac fermions, which originate from the distorted $X$ square net with in-plane lattice polarization.
We found that BaMnBi$_2$ has approximately one-tenth the lattice distortion of BaMnSb$_2$, from which a different configuration of spin-polarized Dirac valleys is theoretically predicted.
This was experimentally observed as a clear difference in the Shubnikov-de Haas oscillation at high fields between the two materials.
The chemically tunable spin-valley coupling in BaMn$X_2$ makes it a promising material for various spin-valleytronic devices.
\end{abstract}

\newpage

\section*{Introduction}
In crystals with broken inversion symmetry, the spin-orbit coupling (SOC) results in a spin- and momentum-dependent splitting of the energy bands. 
In 2D electron gases with structural inversion asymmetry, this leads to spin-momentum locking on the Fermi surface whose type is determined by the relative direction between the 2D plane and the internal electric field due to the asymmetric potential gradient. 
At surfaces and interfaces of heterostructures, for instance, the electric field is generated perpendicular to the 2D plane. This results in Rashba-type spin splitting and leads to a momentum-dependent in-plane helical spin polarization\cite{Rashba_review, Rashba_interface, Rashba_Au_1, Rashba_Au_2}.
More recently, giant Rashba-type spin splitting was also found in bulk materials, such as the layered polar semiconductors BiTeI\cite{BTI_ARPES} and GeTe\cite{GeTe_Rashba}, in which novel transport\cite{BTI_nonreci} and optical properties\cite{BTI_opto} associated with spin-momentum locking have been experimentally observed.

\par

In 2D systems with in-plane inversion asymmetry, on the other hand, an out-of-plane Zeeman-like field is induced by the SOC, the sign of which switches depending on the position in momentum space. 
When the Fermi pockets (i.\,e., electronic valleys) are centered at low-symmetry points, spin-valley coupling occurs and provides a platform for exploring the novel spin physics and spintronic applications associated with the valley degree of freedom. 
A promising system for this is the family of monolayer transition metal dichalcogenides (TMDCs) such as MoS$_2$, which has been attracting significant interest. 
The trigonal prismatic layer of a TMDC consists of a transition metal plane sandwiched between two chalcogen planes\cite{JAWilson_TMDC}. The prismatic layer is regarded as a graphene-like material with different A- and B-site atoms that break the inversion symmetry. 
Therefore, the electron and hole valleys located at the K and K' points are well described by massive Dirac fermions exhibiting valley-contrasting spin polarization perpendicular to the 2D plane\cite{MoS2_valleytro1, MoS2_band_calc, MoS2_QO, MoS2_Xiao_PRL}.
Such spin-valley coupling has indeed been demonstrated through its distinctive optical and transport properties, such as valley-dependent circular dichroic photoluminescence\cite{MoS2_valleytro2, MoS2_monolayer_nanotech,MoS2_monoleyer_natcommun} and nonreciprocal charge transport\cite{MoS2_nonreci}.

\par

To further explore these novel physical properties, it is necessary to control a variety of spin-valley-coupled states.
For this, it is important to realize the states in the bulk form of the TMDCs as well as in the thin film form. In the former, chemical substitution can be used to tune the physical parameters determining the spin-valley coupling such as the magnitudes of the SOC and internal electric field.
However, the spin-valley coupling is suppressed in the most common bulk phase of MoS$_2$ (so-called 2H polytype) because the inversion symmetry is restored\cite{JAWilson_TMDC}. 
Instead, another polytype (so-called 3R) with a non-centrosymmetric layer stacking needs to be stabilized by carefully optimizing the crystal growth\cite{3R_MoS2}. 
Furthermore, in principle, the valley positions in the TMDCs are fixed at the corners of the hexagonal Brillouin zone (the K and K' points)\cite{MoS2_band_calc, MoS2_Xiao_PRL}, reflecting the graphene-like structure.
Such structural restrictions may hamper the wide-range control of the spin-valley coupling in TMDC-related materials.

\par

Recently, the layered Dirac material BaMnSb$_2$ was found to have a polar structure with broken in-plane inversion symmetry in the bulk form\cite{BMS_HS, BMS_Mao}.
BaMnSb$_2$ belongs to the $A$Mn$X_2$ ($A$:\,alkaline earth, $X$:\,Bi, Sb) series of compounds \cite{SMB_Joonbum, SMB_Wang_2, EMB_AFMay, EMB_Masuda_sci_adv, EMB_Kondo, BMS_JLiu_sci_rep, BMS_PNAS}, in which each compound consists of a $A^{2+}$-Mn$^{2+}$-$X^{3-}$ blocking layer and a $X^-$ Dirac fermion layer (see Fig.\,1a).
Since the former layer is an antiferromagnetic insulator with a N\'{e}el temperature around room temperature\cite{SMB_Joonbum, SMB_Wang_2, EMB_AFMay, BMS_JLiu_sci_rep, BMS_PNAS, SMB_neutron_YFGuo, EMB_masuda_sakai_neutron, EMB_neutron}, the quasi-2D Dirac fermion states are formed in the bulk form.
For the layer stacking with a coincident arrangement of the $A$-sites above and below the $X^-$ layer, the $X^-$ layer tends to form a square net, leading to a non-polar tetragonal structure\cite{SMB_Joonbum, SMB_Wang_2, EMB_AFMay, EMB_Masuda_sci_adv, EMB_Kondo}.
However, in BaMnSb$_2$, the Sb$^-$ layer is slightly distorted to an orthorhombic one with zig-zag chains, leading to lattice polarization along the plane. 
Together with the SOC of Sb, this results in the formation of two Dirac valleys with valley-contrasting out-of-plane spin polarizations around the Y point, as was revealed by first-principles calculation and angle-resolved photoemission spectroscopy\cite{BMS_HS}.

\par

Taking advantage of the bulk layered structure, it is anticipated that the spin-valley-coupled Dirac fermion states in BaMnSb$_2$ can be widely tuned by chemical substitution.
In particular, control of the SOC magnitude is most important as the primary determinant of the spin-valley coupling.
For this purpose, the substitution of Sb with Bi in the Dirac fermion layer is a promising approach. 
However, it has been reported in the literature\cite{BMB_powder} that BaMnBi$_2$ has a non-polar tetragonal structure ($I4/mmm$) on the basis of powder X-ray diffraction.
In contrast, by performing synchrotron X-ray diffraction and optical measurements on BaMnBi$_2$ single crystals, we find here that BaMnBi$_2$ has a polar orthorhombic structure similar to that of BaMnSb$_2$.
Reflecting the large difference in the orthorhombicity and SOC between the two materials, our first-principles calculation predicts that the configurations of the spin-polarized Dirac valleys are totally different between the two materials; BaMnBi$_2$ has multiple valleys around the X and Y points.
This is consistent with the complicated behavior of the Shubnikov-de Haas (SdH) oscillation observed at high fields in BaMnBi$_2$.

\section*{Results}
\subsection{Polar crystal structure}
The crystal structure of BaMnBi$_2$ (Fig.\,1a) was determined by single-crystal X-ray structural analysis (for details see Methods, Supplementary Methods and Supplementary Table 1).
The obtained result shows that BaMnBi$_2$ has a polar orthorhombic structure with the same space group as BaMnSb$_2$ (\textit{Imm2}), whereas the orthorhombicity $(c-a)/a$ is estimated to be $\sim0.15\%$, which is only $1/10$ that of BaMnSb$_2$ ($\sim1.3\%$)\cite{BMS_HS}.
Despite such a small orthorhombicity, twin domains are clearly observed with a polarized microscope at room temperature, as shown in Fig.\,1b.
Therefore, the polar structure is stable irrespective of the antiferromagnetic order of Mn sublattice, which sets in at $T_N=290$ K\cite{BMB_powder}.
Among the $X=$ Bi compounds, all of which are known to have the Bi square-net structure\cite{SMB_Joonbum, EMB_AFMay}, BaMnBi$_2$ is the only exception that has Bi zig-zag chains accompanied by in-plane lattice polarization along the $c$-axis.

\par

To verify the polar lattice structure directly, we measured the optical second harmonic generation (SHG), i.e., the frequency doubling of the probing light wave (Fig.\,1c), in BaMnBi$_2$ single crystals at room temperature.
SHG is particularly useful for detecting polar states because, to leading order, SHG occurs only in non-centrosymmetric media. 
Fig.\,1d shows the rotational anisotropy of the SHG intensity from a nearly single-domain region on a cleaved surface.
The rotational anisotropy was obtained by projecting the component of the SHG light oriented parallel to the polarization $\phi$ of the incident fundamental light while $\phi$ was rotated over $360^{\circ}$.
A clear two-fold rotation anisotropy with peaks at $\phi\sim90^{\circ}, 270^{\circ}$ was observed despite the small orthorhombicity.
The data is well fitted by assuming the $mm2$ symmetry with in-plane lattice polarization along the $c$-axis (solid curve in Fig.\,1d, see Supplementary Methods and Supplementary Fig.\,1 for details), which is consistent with the results of the structural analysis.

\par

Similar optical properties associated with polar lattice distortions have also been observed in BaMnSb$_2$\cite{BMS_Mao}.
However, it should be noted that the orthorhombicity of BaMnBi$_2$ is much smaller than that of BaMnSb$_2$ ($\sim1/10$) while the SOC of Bi is stronger than that of Sb.
Because the spin-splitting of the Dirac bands depends on the magnitudes of the polar lattice distortion and the SOC, the spin-valley coupling may significantly differ between these two materials.
To demonstrate this, we calculated the band structure of BaMnBi$_2$ based on the experimentally obtained crystal structure.

\par

\subsection{Electronic structure}
The calculated band structure of BaMnBi$_2$ is shown in Fig.\,2a.
Two sharp Dirac-like bands accompanied by significant spin splitting are formed near the X and Y points (shaded regions in Fig.\,2a).
We show in Fig.\,2b, the spin-resolved Fermi surfaces at the Fermi energy $\epsilon_F=0$ eV. 
All the Fermi surfaces are 2D cylindrical, consistent with the large resistivity anisotropy $\rho_{zz}/\rho_{xx}\sim 100$ observed over the measured temperatures (see Supplementary Fig.\,2), where $\rho_{xx}$ and $\rho_{zz}$ are the in-plane and interlayer resistivities, respectively.
Dirac bands form two- and four-electron valleys (named $\beta$ and $\alpha$) around the X and Y points, respectively.
Each valley is spin-polarized because one of the spin-split Dirac bands crosses $\epsilon_F$.
Because of the lattice polarization along the $c$-axis, the spin polarization of $s_z$ switches sign with respect to the $\Gamma$-Y line\cite{Nagaosa_Iwasa_Zeeman_spinsplitting}; up-spin (red) and down-spin (blue) occur $k_x>0$ and $k_x<0$, respectively.
This $k_x$-dependent $s_z$ polarization (so called Zeeman-type spin splitting) is qualitatively explained by the SOC Hamiltonian $H_{SOC}\propto\bm{\sigma}\cdot(\bm{k}\times\bm{E})=\sigma_z k_xE_y$, where $\bm{\sigma}=(\sigma_x, \sigma_y, \sigma_z)$ is the vector of spin Pauli matrices and $\bm{E}=(0, E_y, 0)$ is the internal electric field.
In Figs.\,2d and f, we show the detailed band dispersion along the $k_y=0.8\pi/c$ and the $k_y=0$ lines that are located at the centers of the $\alpha$ and $\beta$ valleys, respectively. 
In both valleys, the magnitude of the spin-splitting (typically 200 - 300 meV as indicated by vertical arrows) is much larger than the energy gap at the Dirac points ($\sim50$ meV), resulting in spin-polarized nearly-massless Dirac bands.
From the calculation, it is predicted that hole valleys also appear on the $\Gamma$-M line in a similar fashion to the Bi square-net analogs\cite{SMB_Feng_ARPES, EMB_Borisenko, SMB_TBtheory}.
However, in BaMnBi$_2$, the corresponding bands have a large band gap ($\sim250$ meV) due to the lattice distortion (Fig.\,2e).
Furthermore, in experiments, the Hall resistivity $\rho_{yx}$ has a negative slope with respect to the field, indicating that electron-type carriers are dominant in BaMnBi$_2$
(Fig.\,3c).
Therefore, it is plausible that the valence bands along the $\Gamma$-M line do not cross $\epsilon_F$ in reality.
Thus, the obtained Dirac valleys ($\alpha$ and $\beta$) for BaMnBi$_2$ are totally different from those for BaMnSb$_2$; the latter hosts only two equivalent valleys near the Y point along each M-Y line\cite{BMS_HS, BMS_Mao}.
This suggests that the spin-valley coupling of the Dirac fermions in BaMn$X_2$ is highly sensitive to the lattice distortion and the SOC.

\par

\subsection{Quantum oscillation reflecting spin-polarized Dirac fermions}
To reveal this unique spin-valley coupling through the transport properties, we performed magnetoresistivity measurements with pulsed high magnetic fields of up to $\sim$ 56 T.
$\rho_{zz}$ (Fig.\,3a) and $\rho_{xx}$ (Fig.\,3b) exhibit positive magnetoresistance, 
on which clear SdH oscillations are superimposed at low temperatures (below $\sim70$ K).
To clarify the Landau level (LL) structure, we here analyze the SdH oscillation in the conductivity tensors.
In the inset to Fig.\,3d, we plot the in-plane conductivity $\sigma_{xx}=\rho_{xx}/(\rho_{xx}^2+\rho_{yx}^2)$ and interlayer conductivity $\sigma_{zz}=1/\rho_{zz}$ as a function of $B_F/B$, where $B_F$ is the SdH frequency and $B_F/B$ corresponds to the filling factor normalized by the spin-valley degeneracy factor\cite{BMS_HS, Graphite_QO}.
Importantly, the SdH oscillations of $\sigma_{xx}$ and $\sigma_{zz}$ resemble each other well, including the fine structures at high fields ($B_F/B<0.5$).
Therefore, both SdH oscillations directly reflect the density of states in the LLs.
In the following, we use $\sigma_{zz}$ for the detailed analysis owing to the higher S/N ratio and clearer oscillations even at low fields.

\par

In Fig.\,3d, we plot $-d^2\sigma_{zz}/d(1/B)^2$ vs $1/B$, where the oscillation nearly consists of a single frequency at $1/B\geq0.05$ T$^{-1}$.
From the Lifshitz-Kosevich equation, the oscillation component is given by\cite{LK_eq, YAndo_review}
\begin{equation}
\Delta\sigma\propto\cos\left\{2\pi\left(\frac{B_F}{B}-\frac{1}{2}+\frac{\gamma}{2\pi}\right)\right\},
\end{equation}
where $\gamma$ is the Berry phase.
Following this equation, we obtained the linear fan diagram (Fig.\,3e) by indexing the integer (half-integer) $N$ to the peaks (dips).
The gradient corresponds to the frequency $B_{F1}$ ($=12.9$ T).
The vanishing $N$-intercept ($=0.01$) indicates the non-trivial Berry phase $\gamma=\pi$ consistent with the nearly massless Dirac bands (Figs.\,2d, f).
Note that at $1/B<0.05$ T$^{-1}$ corresponding to the zeroth LL, additional structures (vertical arrows in Fig.\,3d) appear, which cannot be explained by the extrapolation of the low-field LLs ($N>1$).
We will discuss the origin of these complicated LLs in the quantum limit later.

\par

The application of magnetic fields should lift the degeneracy of the equivalent valleys with opposite spin polarizations because of the Zeeman energy $\epsilon_Z=g^*\mu_BB$, where $g^*$ is the effective $g$ factor and $\mu_B$ is the Bohr magneton.
However, no clear splitting is observed in the SdH oscillation for $N\geq1$ (Fig.\,3d).
In 2D systems, the ratio of $\epsilon_Z$ to the cyclotron energy $\epsilon_c=e\hbar B\cos\theta/m_c$ is given by $\epsilon_Z/\epsilon_c=g^*(m_c/2m_0)\cos\theta$, where $m_c$ is the cyclotron mass, $m_0$ is the bare electron mass, and $\theta$ is the tilt angle from the normal to the 2D plane. 
This shows that the spin splitting is enhanced by increasing $\theta$, which leads to the variation of the amplitude and phase of the SdH oscillation with $\theta$.
However, this is not the case in BaMnBi$_2$.
Figure 4 shows $\sigma_{zz}$ plotted against $1/B\cos\theta$. The oscillation period is almost constant for all values of $\theta$ up to 77$^{\circ}$ (as shown by vertical dotted lines), reflecting the cylindrical quasi-2D Dirac valleys.
In addition, the amplitude and phase of the SdH oscillation are almost unchanged upon $\theta$ without any enhancement of the splitting, indicating that $\epsilon_z/\epsilon_c$ independent of $\theta$, i.\,e.\,$\epsilon_z\propto B\cos\theta$ (not $\propto B$).
This reflects that the spins of the Dirac bands are almost fixed along the $s_z$ direction even under the tilted fields owing to the SOC, providing firm evidence for the spin-valley coupling.
For BaMnBi$_2$, as shown in Figs.\,2d and 2f, the magnitude of the SOC-induced spin splitting is theoretically estimated to be 200 - 300 meV at each valley.
This corresponds to the effective Zeeman field of 1700 - 2500 T, which is much higher than the applied field.
The similar tilt angle dependence has also been observed in other spin-valley-coupled systems, such as BaMnSb$_2$\cite{BMS_HS} and monolayer MoS$_2$\cite{MoS2_QO}

\par

\section*{Discussion}
In spite of the small orthorhombicity of $(c-a)/a\sim0.15\%$ in BaMnBi$_2$, it is surprising that the Dirac bands around the X and Y points show the large spin splitting of 200 - 300 meV (Figs.\,2d and f), which is comparable to that for BaMnSb$_2$\cite{BMS_HS, BMS_Mao}.
We estimate the magnitude of the in-plane lattice polarization by calculating the off-center displacement of the Bi$^-$ layer relative to the Ba$^{2+}$ layer.
Note that the off-center displacement of the [MnBi]$^-$ layer relative to the Ba$^{2+}$ layer is negligibly small compared to that of the Bi$^-$ layer.
The unit cell of the Bi$^-$ layer together with the Ba$^{2+}$ ion is shown in Fig.\,2c. Two inequivalent Bi$^-$ ions (Bi1 and Bi2) exist because of the zig-zag chain structure.
From the structural analysis, the center position of Bi1 ($\bm{G}_{\rm{Bi1}}$) is  estimated to be $\bm{G}_{\rm{Bi1}}=(0, -0.008(3)c)$ by taking the Ba$^{2+}$ ion as the origin, while the center position of Bi2 $\bm{G}_{\rm{Bi2}}=(0, 0.029(3)c)$.
The total displacement of the Bi$^-$ layer is given by $\bm{G}_{\rm{Bi}}=(\bm{G}_{\rm{Bi1}}+\bm{G}_{\rm{Bi2}})/2=(0, 0.011(4)c)$, and reaches $\sim1\%$ of the $c$-axis length.
This value is roughly an order of magnitude larger than the orthorhombicity.

\par

We also estimate the magnitude of the in-plane lattice polarization in BaMnSb$_2$, which has a larger orthorhombicity of $(c-a)/a\sim1.3\%$\cite{BMS_HS}.
From the structural data in Ref.\,\onlinecite{BMS_HS}, it is estimated that $\bm{G}_{\rm{Sb1}}=(0, -0.0082(2)c)$ and $\bm{G}_{\rm{Sb2}}=(0, 0.0693(3)c)$.
Note here that $|\bm{G}_{\rm{Sb1}}|$ is similar to $|\bm{G}_{\rm{Bi1}}|$, while $|\bm{G}_{\rm{Sb2}}|$ is approximately 2.4 times as large as $|\bm{G}_{\rm{Bi2}}|$, which suggests that the magnitude of the lattice polarization is determined by the displacement of Bi2 and Sb2.
The resultant total displacement of the Sb$^-$ layer is given by $\bm{G}_{\rm{Sb}}=(\bm{G}_{\rm{Sb1}}+\bm{G}_{\rm{Sb2}})/2=(0, 0.0305(3)c)$. 
Compared to BaMnBi$_2$, although the orthorhombicity differs by approximately an order of magnitude, the off-center displacement differs only by approximately three times. 
Considering the stronger SOC in Bi$^-$ than in Sb$^-$, it is natural that the spin splitting of the Dirac bands in BaMnBi$_2$ is comparable to that in BaMnSb$_2$.
Nevertheless, the valley configuration sensitively depends on each parameter of the lattice polarization and SOC and hence is totally different between the two compounds.

\par

Finally, we discuss the anomaly in the SdH oscillation at $1/B<0.05$ T$^{-1}$ (Fig.\,3d).
As denoted by triangles in Fig.\,4, the position ($1/B\cos\theta$ value) of the additional structure in the SdH oscillation is unchanged by $\theta$, suggesting it is related to the cyclotron motion.
Furthermore, because such an additional oscillation has not been reported in BaMnSb$_2$\cite{BMS_HS, BMS_Mao}, it is characteristic of BaMnBi$_2$. 
Considering that BaMnBi$_2$ has two inequivalent Dirac valleys ($\alpha$ and $\beta$ in Fig.\,2b), the additional oscillation may arise from another SdH oscillation superimposed on the main one ($B_{F1}=12.9$ T, see Fig.\,3e). 
By the analysis based on this assumption, we obtain the frequency of $B_{F2}=61(2)$ T from the oscillatory component at high fields ($1/B<0.05$ T$^{-1}$, see Supplementary Note, Supplementary Figs.\,4a-d and 5, and Supplementary Table 2). 
Importantly, the Fermi surfaces corresponding to $B_{F1}$ and $B_{F2}$ are semi-quantitatively reproduced by the first-principles calculation taking account of a slight shift in $\epsilon_F$ (by $\sim$30 meV, see Supplementary Fig.\,4e). 
We also note that such multiple carriers in BaMnBi$_2$ are also consistent with the non-linear $\rho_{yx}$ at low fields (Fig.\,3c and Supplementary Fig.\,3). 
For these reasons, it is likely that the anomaly at $1/B<0.05$ T$^{-1}$ originates from the SdH oscillation of another Dirac valley. 
However, to clarify the precise origin, further investigations on the Fermi surface, such as photoemission spectroscopy, will be necessary as a future work.

\par

In conclusion, we have revealed the crystal and electronic structure of the layered Dirac material BaMnBi$_2$.
From the single-crystal X-ray diffraction and the SHG measurement, we found that the Bi$^-$ square-net is slightly distorted to form a zig-zag chainlike structure with in-plane lattice polarization.
The first-principles calculation predicts that as a result of the broken inversion symmetry together with the SOC, the Dirac valleys show valley-contrasting spin polarization along the $s_z$ direction.
Such spin-valley coupling is indeed supported by the peculiar dependence of the SdH oscillation on the tilt angle of the field.
Interestingly, it is also predicted that the valley configuration in BaMnBi$_2$ (four and two valleys around the X and Y points, respectively) is totally different from that in the Sb analog BaMnSb$_2$ (two valleys around the Y point).
This difference originates from the differing magnitudes of the lattice polarization and the SOC; the experimentally obtained lattice distortion in BaMnBi$_2$ is $\sim1/10$ of that in BaMnSb$_2$, while the SOC in Bi is larger than that in Sb.
Our results demonstrate that the spin-valley coupling of Dirac fermions in BaMn$X_2$ is finely tunable by chemical substitution and is promising for a variety of novel applications, including electronic and optoelectronic devices utilizing the spin and valley degrees of freedom. 
Furthermore, the high sensitivity of BaMn$X_2$ to lattice distortions may be taken advantage of to control spin-valley-coupled phenomena using mechanical stimuli, such as strain\cite{BMS_piezo}.

\begin{methods}
\subsection{Crystal growth}
Single crystals of BaMnBi$_2$ were grown by the Bi self-flux method.
99.99\% Ba, 99.9\% Mn, and 99.999\% Bi metal ingots were mixed in an alumina or carbon crucible at a ratio of Ba$:$Mn$:$Bi$=1:1:4$ and sealed in an evacuated quartz tube. 
The ampoule was heated at 1000$^{\circ}$C for 1 day and cooled down to 400$^{\circ}$C at a 2$^{\circ}$C/h cooling rate.
After the ampoule was turned upside down, the flux and single crystals were separated by centrifugation.
The synthesized crystals have a plate-like shape with the typical size of 2 mm$\times$2 mm$\times$1 mm.

\subsection{Single crystal X-ray diffraction}
A block-shaped pale black single crystal with the dimensions of 0.15$\times$0.15$\times$0.12 mm was mounted on a loop and set on a RIGAKU 1/4chi goniometer with PILATUS3 X CdTe 1M in SPring-8 BL02B1. 
The diffraction data were collected using synchrotron radiation ($\lambda=$ 0.4121 \AA) monochromated by Si(311) at $T=100$ K. 
The diffraction data from 7570 data points within $3.954^{\circ}\leq2\theta\leq59.298^{\circ}$ were collected and merged to give 3236 unique reflections with the $R_{int}$ of 0.049.
The structure was solved by a dual-space method and refined on $F^2$ by a least-squares method using the programs SHELXS\cite{CS1} and SHELXL-2018/3\cite{CS2}, respectively. 
The anisotropic atomic displacement parameters were applied for all atoms.
The final $R$ values from 3236 unique reflections ($2\theta_{max} = 59.298^{\circ}$) with $I > 2\sigma(I)$ are 0.0659 and 0.1759 for $R(F)$ and $wR(F^{2})$, respectively. 
The Flack parameter ($\chi$=0.39(2))\cite{CS3, CS4} from anomalous scattering suggests that two types of absolute structures ($+P$ and $-P$) are mixed at a ratio of 0.39:0.61.
For further details, see the Supplementary information.

\subsection{Optical measurements}
Optical second harmonic generation (SHG) is a nonlinear optical process in which two photons with a frequency of $\omega$ interact with the material and produce a photon with the frequency of $2\omega$.
Under the electric dipole (ED) approximation, the SHG is given by $P_i(2\omega) = \varepsilon_0 \sum_{j, k} \chi_{ijk}^{(2)} E_j(\omega) E_k(\omega)$, where $\varepsilon_0$, $E_j(\omega)$, $E_k(\omega)$, and $P_i(2\omega)$ are the permittivity of vacuum, the $j$- and $k$-polarized incident electric fields, and the induced $i$-polarized nonlinear polarization, respectively.
The ED-type SHG is only allowed in noncentrosymmetric materials or surfaces/interfaces at which the inversion symmetry is broken.
The nonlinear susceptibility $\chi_{ijk}^{(2)}$ is sensitive to the symmetry of the materials and its nonzero components were determined by the Neumann principle\cite{Becher2015, Matsubara2010}.
A cleaved surface of a BaMnBi${}_2$ single crystal was irradiated by light pulses from a Ti:Sapphire laser with the wavelength of 800 nm, pulse width of 100 fs, and repetition rate of 80 MHz.
The reflection geometry SHG measurements were at nearly normal incidence ($\sim1^\circ$) with a typical laser power of 40 mW and laser beam spot size of $\sim250\,\mu$m.
All the SHG measurements were performed in vacuum at room temperature.

\subsection{First-principles band calculations}
We performed first-principles band structure calculations using the density functional theory with the generalized gradient approximation\cite{GGA} and the projector augmented wave method\cite{paw} implemented in the Vienna {\it ab initio} simulation package\cite{vasp1,vasp2,vasp3,vasp4}.
The plane-wave cutoff energy of 350 eV and a $12\times 12\times 12$ $\bm{k}$-mesh were used with spin-orbit coupling (SOC) included.
The experimental crystal structure determined by this study was used. 
We assumed the G-type antiferromagnetic order for the Mn atoms.
After the first-principles calculation, we constructed the Wannier functions\cite{Wannier1,Wannier2} using the \textsc{Wannier90} software\cite{Wannier90}.
We did not perform the maximal localization procedure to prevent the mixture of the different spin components.
We took the Mn-$d$ and Bi-$p$ orbitals as the Wannier basis. A $12\times 12\times 12$ $\bm{k}$-mesh was used for the Wannier construction.
By using the tight-binding model consisting of these Wannier functions, we obtained the band structure and the Fermi surface colored with the spin polarization, $\langle s_z \rangle$.
The Fermi surface was depicted with a $140\times 140\times 140$ $\bm{k}$-mesh using the \textsc{FermiSurfer} software\cite{FSurfer}.

\subsection{Transport measurements}
The temperature dependence of $\rho_{xx}$ and $\rho_{zz}$ was measured by a conventional 4-probe DC method using a Physical Property Measurement System (Quantum Design).
The field dependence of $\rho_{xx}$, $\rho_{yx}$, and $\rho_{zz}$ up to $\sim56$ T was measured using the non-destructive mid-pulse magnet at the International MegaGauss Science Laboratory at the Institute for Solid State Physics, University of Tokyo.
In this measurement, we used a 4-probe AC method with a typical current and frequency of 5 mA and 50 kHz, respectively.
The typical sample sizes used in the in-plane resistivity and interlayer resistivity measurements were 1 mm$\times$0.3 mm$\times$0.1 mm (S\#3) and 0.8 mm$\times$0.3 mm$\times$0.2 mm (S\#1, S\#2), respectively.
The transport properties are almost the same among the measured samples (see Supplementary Fig.\,6 and Supplementary Table 3).
\end{methods}

\section*{Data Availability}
The data that support the findings of this study are available from the corresponding authors upon reasonable request.

\begin{addendum}
\item We are grateful to M.\,Hagiwara and Y.\,Shimizu for fruitful discussions.
We also thank H.\,Fujimura and K.\,Nakagawa for their experimental support.
This work was partly supported by the JST PRESTO (Grant No.\,JPMJPR16R2), the JSPS KAKENHI (Grant Nos.\,19H01851, 19K21851, 19H05173, 21H00147, 17H04844, 21H04649 and 18H04226), and the IWATANI NAOJI Foundation.
The synchrotron radiation experiments were performed at the BL02B1 of SPring-8 with the approval of the Japan Synchrotron Radiation Research Institute (JASRI) (Proposal No. 2019B1402). 
\item[Competing Interests] The authors declare that they have no
competing interests.
\item[Correspondence] Correspondence and requests for materials
should be addressed to H.\,S. and M.\,K.
\item[Author contributions] H. S. conceived the project and planned the experiments. 
M. K. and H. S. worked on the single crystal growth. 
T. K, M. K., and H. S. measured the single-crystal x-ray diffraction and T. K analyzed the data. 
D. S. and M. M. performed the optical measurements.
M. O. and K. K performed the first-principles calculations. 
M. K., H. S., R. K., A. M., and M. T. performed the high-field transport measurements. 
M. K., H. S., H. M., and N. H. measured the resistivity at low fields. 
M. K. and H. S. wrote the manuscript with the inputs from M. O., T. K., D. S., and M. M. 
All the authors discussed the results and commented on the manuscript.
\end{addendum}

\newpage

\begin{figure}
\begin{center}
\includegraphics[width=14cm]{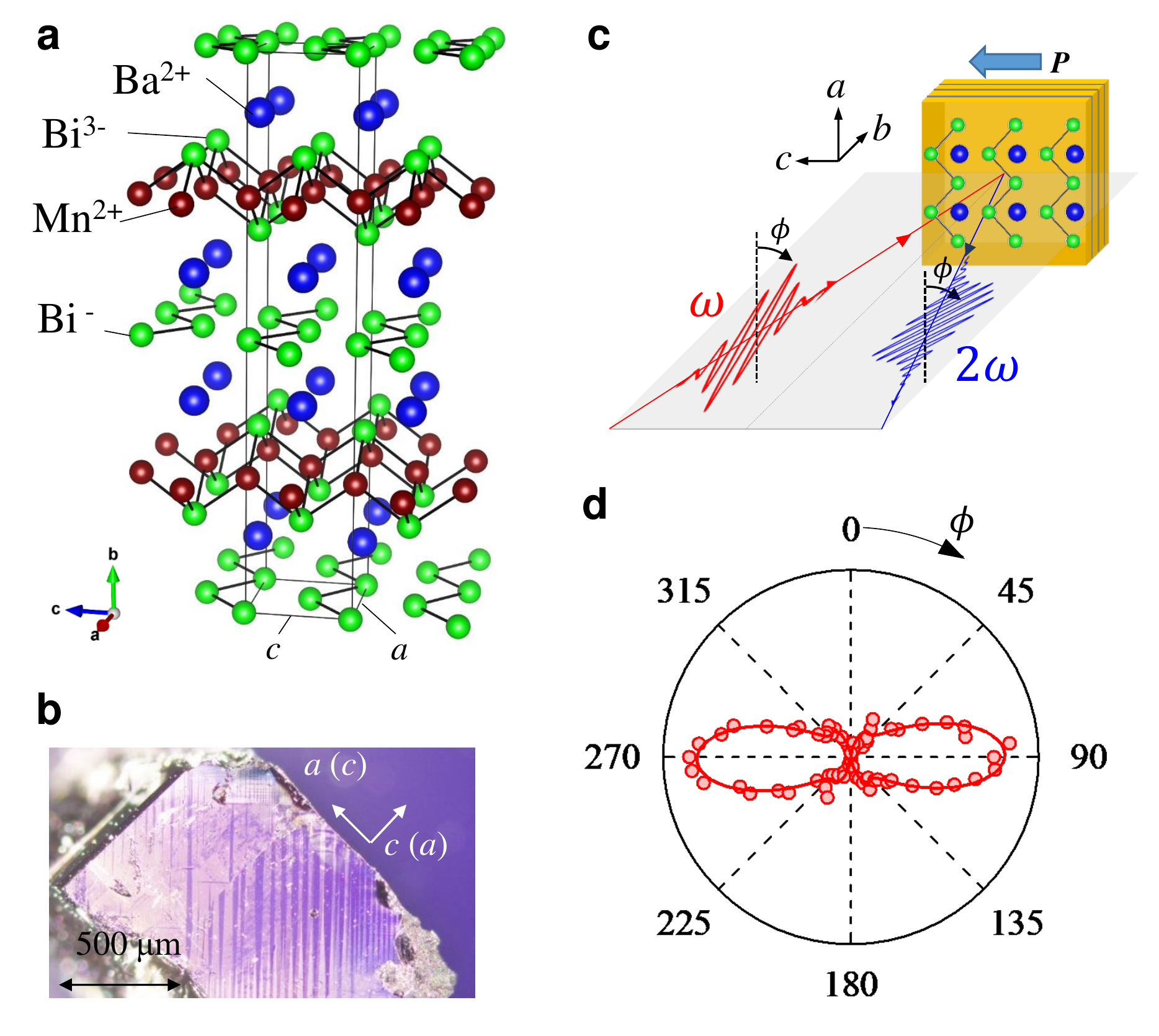}
\end{center}
\caption{
\textbf{Polar crystal structure and optical properties of BaMnBi$_2$.}
\textbf{a}, Crystal structure determined from single-crystal X-ray diffraction at 100 K.
\textbf{b}, The polarized microscope image of the as-grown crystal surface.
\textbf{c}, Schematic image of the SHG measurement at nearly normal incidence ($\sim1^{\circ}$).
The angle $\phi$ denotes the polarization of the incident fundamental light ($\omega$) and the emitted SHG light ($2\omega$).
\textbf{d}, Polarization analysis of the SHG signal with a fitting (red solid curve) assuming $mm2$ symmetry.
All the measurements were performed at room temperature.
}
\end{figure}

\newpage

\begin{figure}
\begin{center}
\includegraphics[width=17cm, clip]{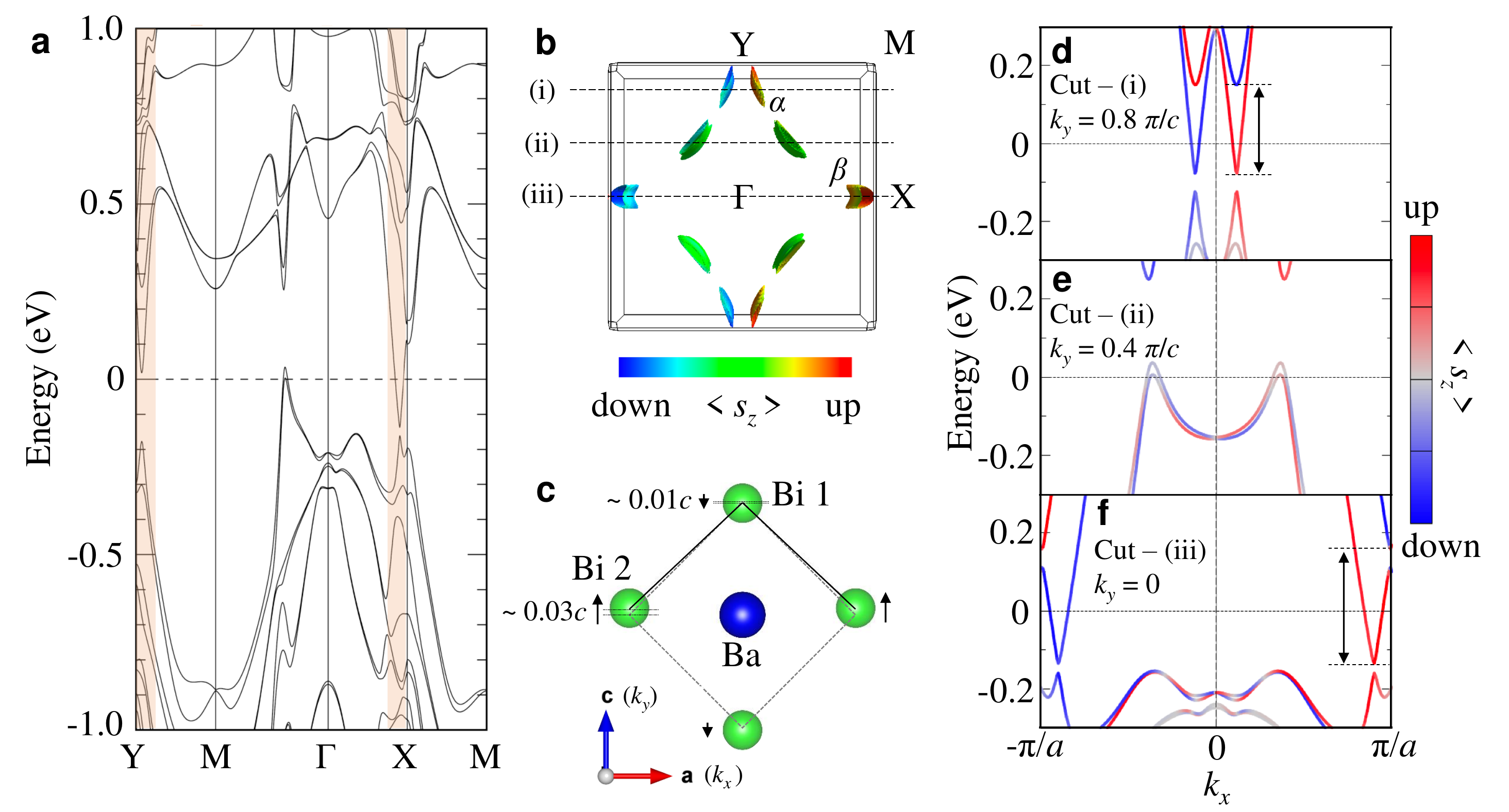}
\end{center}
\caption{
\textbf{Band structures and spin-polarized Fermi surfaces.}
\textbf{a}, Calculated band structure of BaMnBi$_2$ with SOC.
\textbf{b}, Spin-resolved Fermi surfaces calculated at $\epsilon_F=0$ eV.
The color of Fermi surfaces represents the spin polarization $\left<s_z\right>$.
\textbf{c}, Atomic displacements (denoted by arrows) within the Bi$^-$ layer deduced from the structural analysis.
Owing to the lattice distortion, the black solid lines connecting the centers of Bi1 and Bi2 deviate from the gray dotted lines that describe a square centered at Ba.
\textbf{d-f}, Energy dispersion cuts along the dashed lines (i)-(iii) shown in \textbf{b}.
The red and blue colors represent spin up and down, respectively. 
Vertical double-headed arrows in \textbf{d} and \textbf{f} denote the spin splitting energy.
}
\end{figure}

\newpage

\begin{figure}
\begin{center}
\includegraphics[width=17cm]{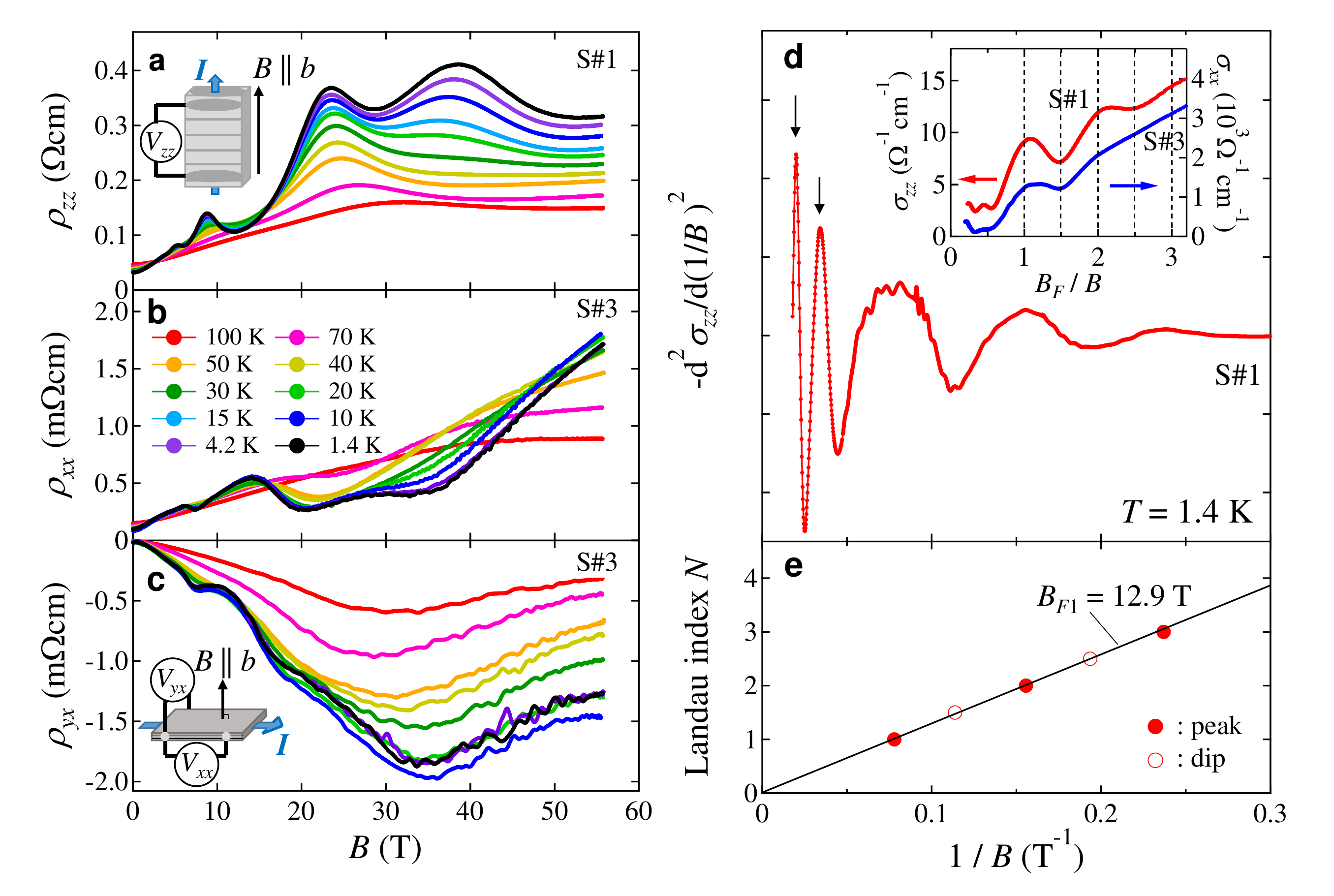}
\end{center}
\caption{
\textbf{Magnetoresistivity at high fields and analysis of the quantum oscillations.}
\textbf{a}, \textbf{b}, \textbf{c}, Field ($B$) dependence of $\rho_{zz}$ (\textbf{a}), $\rho_{xx}$ (\textbf{b}), and the in-plane Hall resistivity $\rho_{yx}$ (\textbf{c}) of BaMnBi$_2$.
The insets in \textbf{a} and \textbf{c} illustrate the configurations for the interlayer and in-plane measurements.
\textbf{d}, $-d^2\sigma_{zz}/d(1/B)^2$ at $T = 1.4$ K plotted as a function of $1/B$.
The inset shows $\sigma_{zz}$ (red) and $\sigma_{xx}$ (blue) as a function of the normalized filling factor $B_F/B$.
\textbf{e}, The Landau fan diagram obtained from $-d^2\sigma_{zz}/d(1/B)^2$.
The black solid line is the linear fit to the experimental data.
}
\end{figure}

\newpage

\begin{figure}
\begin{center}
\includegraphics[width=12cm]{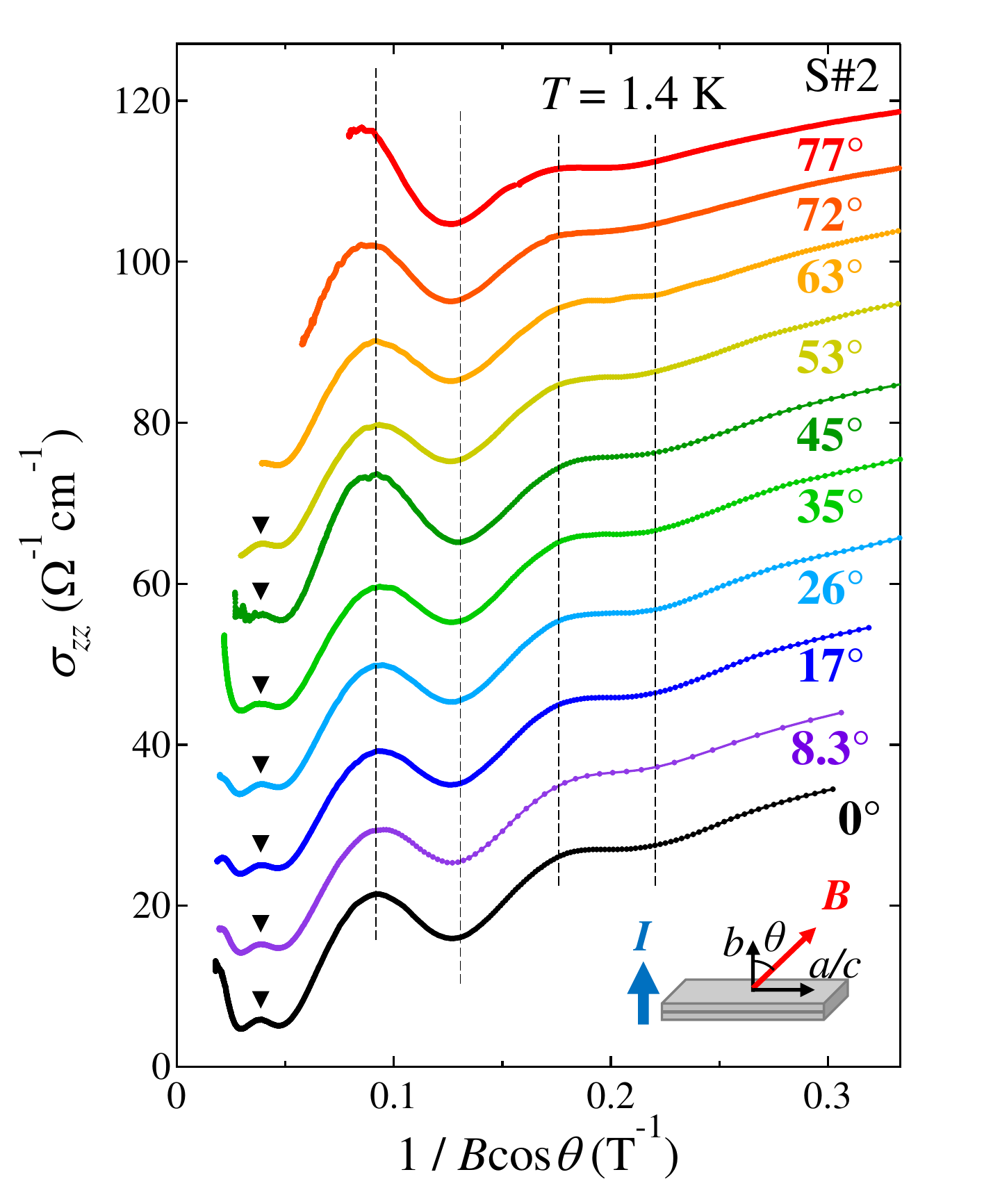}
\end{center}
\caption{
\textbf{Angular dependence of quantum oscillation.}
The $\sigma_{zz}$ data measured at various field tilt angles ($\theta$) are plotted versus $1/B\cos\theta$, where $\theta$ is the angle between the $b$-axis and the field (inset).
Each curve at $\theta\geq8.3^{\circ}$ is shifted vertically (by 10 $\Omega^{-1}$cm$^{-1}$)
The vertical dashed lines denote the positions of the $\sigma_{zz}$ peaks and dips arising from the SdH oscillation.
The triangles denote the positions of the additional peak structure at high fields.
}
\end{figure}

\end{document}